\documentclass{ws-procs975x65}

\begin{document}
\begin{flushright}
Preprint INR-TH-2015-024
\end{flushright}

\markboth{Grigorii B. Pivovarov}{Generization of non-abelian gauge theories}

%
%

\title{Gell-Mann--Low scheme for the Standard Model}

\author{Grigorii B. Pivovarov}

\address{Theory division, Institue for Nuclear Research, Prospekt 60letiya Oktyabrya 7a,\\
Moscow, 117312,
Russia\\
gbpivo@ms2.inr.ac.ru}



\begin{abstract}
I describe a version of the Gell-Mann--Low scheme which is applicable to the Standard Model. First, I give a procedure for extracting input parameters of the theory from the Green's functions. After that, an iterative procedure of expressing the Green's functions in terms of the above parameters is given. 
\end{abstract}

\bodymatter

\section{Introduction}

Earlier I argued \cite{Pivovarov:2015tya} that it is highly desirable to develop a version of the Gell-Mann--Low scheme for the Standard Model. Neglecting important details, my argument was that it is not healthy to keep to a unique way of doing something vitally important. In particular, it is not healthy to take as a unique the `t~Hooft-Weinberg's way of computing the evolution of the Standard Model parameters.  (This obvious argument can be pitched higher---`t~Hooft-Weinberg and Gell-Mann--Low schemes may result in qualitatively different evolutions of their respective input parameters 
for the Standard Model. \cite{Pivovarov:2015tya})

I remind the reader that the `t~Hooft-Weinberg scheme was invented as a technical means to overcome the difficulties of applying the Gell-Mann--Low scheme. Notably, Weinberg had given  a warning \cite{Weinberg:1951ss}: mass independent schemes can not be  applied to theories with scalar fields. At least, extra work should be done to make this possible.  

Within a year such extra work had been done \cite{Collins:1973yy}: it was demonstrated that the MS-scheme can be used to treat scalar fields. Since then the `t~Hooft-Weinberg approach in conjunction with the MS-scheme had become the de facto standard for considering the remormalization group evolution of the Standard Model parameters.

The standard approach is elegant and efficient. But applying it to theories with scalar fields may be a source of a trouble. I mean the so-called naturalness problem. 

It was pointed out early \cite{Susskind:1978ms} that using an UV cutoff instead of UV renormaliztion results in large quantum corrections to the scalar masses proportional to the cutoff. Nothing like this takes place within the MS-scheme.

An explanation to this was suggested \cite{Veltman:1980mj}: dimensional regularization for theories with scalar particles yields Feynman integrals contributing to the scalar self-energy with poles in the plane of complex dimension $D$ at $D=4-2/N$, where $N$ is the number of loops in the integral. This is the way the quadratic divergences manifest themselves within the dimensional regularization. The minimal subtractions ignore these poles, because they are away from the physical dimension at finite number of loops. 

This observation does not mean immediately that the MS-scheme is inapplicable to models with scalar fields. Rather, it justifies in my view the need to try a nonstandard approach for describing the renormalization group evolution of the parameters of the 
Standard Model.

I describe below a version of the Gell-Mann--Low scheme applicable in principle to the Standard Model. The description is broken into three steps. 

First, in the next Section, I describe how input parameters of the model can be extracted from the connected Green's functions of the model. After that, in Section 3, I express the bare action of the model in terms of the parameters introduced in Section 2. Finally, I describe the expression of the desired Green's functions in terms of the above parameters. 

The description given below is rather sketchy. I only give basic formulas without derivation. Some of the derivations are given elsewhere \cite{Pivovarov:2009wa,Kim:2014aba}.

\section{Extraction of parameters}

In the most general terms, The Gell-Mann--Low scheme for describing the remormalization group evolution of model parameters consists in the following. The first ingredient is a prescription on extracting the input parameters of the theory from the Green's functions of the theory. The second ingredient is an algorithm of expressing the Green's functions in terms of the input parameters defined previously. The third and the last ingredient is a derivation of equations for Green's functions expressing the independence of Green's functions on the way the input parameters were extracted.

In this Section I concentrate on the extraction of input parameters from the Green's functions. 

It all starts with the propagator matrix:
\begin{equation}
\label{propagators}
D^{\alpha\beta}\equiv \langle\phi^\beta\phi^\alpha\rangle.
\end{equation}
Here $\phi^\alpha$ is a component of the field $\phi$ of the model. Notice the order of the superscripts. It is to account for the presence of the fermi  components (the matrix $D$ is block-diagonal; the bose block is symmetric, the fermi---anti-symmetric). The field is shifted to achieve $\langle \phi\rangle=0$. All the UV counterterms are included, and $D$ is UV finite.

Next I introduce the inverse propagator matrix $R_{\alpha\beta}$. By definition, $RD\equiv 1$. I will also need ``the local part of $R$,'' $R_L$. To define it, I consider a quadratic form of the fields:
\begin{equation}
\label{quadratic}
Q[\phi]\equiv \frac{1}{2}R^T\phi\phi,
\end{equation}
and a linear projector, acting in the space of quadratic forms of the fields, $P_2Q[\phi]\equiv Q_L[\phi]$, where $Q_L$ is a ``local part of the quadratic form $Q$.''
I use above and in the the following the condensed notation: $R\phi\phi\equiv R_{\alpha\beta}\phi^\alpha \phi^\beta$. Notice, that the transposed matrix $R^T$ appears in the definition of $Q_L$. If not the presence of this transposition, the sign of the fermionic part would come out wrong (see below).

The local part of $Q$ is not defined uniquely. What I require from it is that the quadratic part of the bare action of the model would be in the image of $P_2$: $P_2S_{B,2}[\phi]\equiv S_{B,2}[\phi]$, and the dimension of the image of $P_2$ would equal the number of independent parameters in the quadratic part of the bare action.

The local part of the inverse propagator matrix $R_L$ is defined by $Q_L[\phi]\equiv R^T_L\phi\phi/2$. The idea is that the UV infinite parameters in the quadratic part of the bare action can be parametrized with the same number of UV finite parameters in $R_L$.

Now, to make the above consideration more concrete, I give a particular example of a set of projectors $P_2$ satisfying my requirements. The members of this set are parametrized with a set of field configurations $\{\bar{\phi}\}$. The field configurations of this set, $\bar{\phi}_i\in \{\bar{\phi}\}$,  are in one-to-one correspondence with the parameters in the quadratic part of the bare action (which are related to bare field normalizations and masses):
\begin{equation}
\label{b-action}
S_{B,2}[\phi]\equiv x_{B,i}\mathcal{O}^i[\phi]
\end{equation}
Here $\mathcal{O}^i[\phi]$ is the basis of local UV finite quadratic operators appearing in the bare action.

Let the set of field configurations $\{\bar{\phi}\}$ be such that the matrix 
\begin{equation}
\mathcal{O}^i_j(\{\bar{\phi}\})\equiv\mathcal{O}^i[\bar{\phi}_j]
\end{equation}
is invertible:
\begin{equation}
\mathcal{O}^i_j(\{\bar{\phi}\})\bar{\mathcal{O}}^j_k(\{\bar{\phi}\})=\delta^i_k.
\end{equation}

With these notations, I can define 
\begin{equation}
P_2(\{\bar{\phi}\})Q[\phi]\equiv Q[\bar{\phi}_j]\bar{\mathcal{O}}^j_i(\{\bar{\phi}\})\mathcal{O}^i[\phi].
\end{equation} 
It can be checked that this $P_2(\{\bar{\phi}\})$ is indeed a projector, and that it leaves intact the quadratic form $S_{B,2}[\phi]$ of Eq. (\ref{b-action}).

Now I define the generating functional of amplitudes. To this end, I first define the generating functional of Green's functions
\begin{equation}
\label{GF}
\exp(W[J])=N\int\mathcal{D}\phi\exp(-S_B[\phi]+iJ\phi).
\end{equation}
The normalization $N$ is defined by the condition $W(0)=0$, and the expansion of $W$ in the source $J$ starts from the quadratic part: $W[J]=-DJJ/2+\dots$. Notice that the matrix of the quadratic form is the above $D$.

Now I partly subtract from this $W$ the quadratic part and define an auxiliary object $\bar{W}[J]\equiv W[J]+D_LJJ/2$ where $R_LD_L=1$ (that is, $D_L$ is the propagator matrix up to the non-local self-energy corrections).

At last I define the generating functional of the amplitudes in terms of the auxiliary $\bar{W}$:
\begin{equation}
\label{amplitudes}
A[\phi]\equiv -\bar{W}[-iR_L\phi].
\end{equation}

The amplitudes are obtained by partial amputation of the external legs. Indeed, if the propagator matrix would equal the inverse of $R_L$, the amputation would be complete. This corresponds to allowing loop corrections in the external legs of the amplitudes in the definition (\ref{amplitudes}).   

Next I define the generating functional of three-particle amplitudes. It is the part of $A$ cubic in the fields, $A[\phi]=G\phi\phi\phi/6+\dots$ ($G$ is a tensor with three subscripts), and the terms denoted by the dots are not less than the fourth order in the fields. So, the generating functional of the three-particle amplitudes is $T[\phi]\equiv G\phi\phi\phi/6$.

Analogously to defining $Q_L$, I define $T_L$: it is a local part of the three-particle amplitudes
\begin{equation}
\label{t-part}
T_L[\phi]\equiv P_3T[\phi].
\end{equation}
$P_3$ is defined in complete analogy with $P_2$. The only difference is that it acts on cubic functionals of the fields.

The input parameters of the theory are the parameters in $Q_L$ and $T_L$. They are respectively quadratic and cubic local functionals of the fields. The number of input parameters equals the dimension of the image of the projector $P_2+P_3$. It can also be characterized as the number of independent parameters in the part of the bare action quadratic and cubic in the fields. 

What about the quartic couplings in the bare action? They are not independent parameters of the theory. It is a remarkable fact: all quartic couplings of the Standard Model are functions of the cubic and quadratic couplings. For gauge couplings it is well known. For scalar self-couplings it is less well known bur still true: quartic scalar self-couplings can be expressed in terms of the cubic ones.

It will be used in the following that the quartic couplings are functions of the cubic and quadratic ones. In particular, quartic couplings can be expanded in power series of the cubic couplings, and the expansion starts from the quadratic terms in the cubic couplings. 

My next aim is to describe the expansion of the generating functional $A[\phi]$ in powers of the functional $T_L[\phi]$. The terms of the expansion will be defined in terms of $T_L$ and $Q_L$. The first step in this direction is to express the bare action in terms of $A[\phi]$ and $Q_L[\phi]$.

\section{Bare action}

It was noted a long time ago \cite{deDominicis:1967px} that there is a duality between the bare action and the generating functional of Green's functions (the Dominicis-Englert duality).  A variation of this idea leads to a sort of Feynman rules allowing one to express the bare action in terms of $A$ and $Q_L$:
\begin{equation}
\label{DE}
S_B[\phi]=Q_L[\phi]-\log T\exp\big(-A[\phi]\big).
\end{equation}
Here $T$ is the $T$-product:
\begin{equation}
\label{T-prod}
T\equiv \exp\Big(-\frac{1}{2}D_L\delta_\phi\delta_\phi\Big),
\end{equation}
where $\delta_\phi$ is the variational derivative in the field. Notice, also, that namely $Q_L$ defined in (\ref{quadratic}) appears in the right-hand-side. The transposition of $R$ in this definition is needed to obtain the correct sign by the fermionic part of the first term in the right-hand-side of (\ref{DE}).

In the above formula, bare action is represented as a sum of Feynman amplitudes with vertexes $A$ and propagators $D_L$. A derivation has been discussed previously~\cite{Pivovarov:2009wa} (this derivation uses its own notations, but all the ideas needed to derive (\ref{DE}) are present in the paper cited). 

A theory is usually defined in terms of its bare action. The bare action is defined in its turn in terms of its bare couplings. In the next Section I use an alternative approach. (I have been referring \cite{Pivovarov:2009wa} to it as the inaction approach because it avoids any use of the bare action.) The idea is to explicitly use the conditions defining the bare action as conditions on $A$ and $D_L$ avoiding any appearance of UV infinite objects in the formalism. This will be further discussed in the next Section. Now I want to discuss the restrictions on the bare action characterizing the Standard Model.

The key point here is that writing these restrictions is to write conditions on functionals. It leads to introducing functions of functionals. Functions of fields (functionals) is a subject of a separate treatment known as  \textit{functional methods}~ \cite{Vasiliev:1998cq}. Functions of functionals mapping them to other functionals is probably deserving a separate consideration, but this is not a subject for this text. I have already used two linear functions of this sort---the projectors
$P_2$ and $P_3$. Now I need a nonlinear function of this sort. For this I need to introduce the following notation:
$\lambda[[F]]$ will denote a function $\lambda$ of the functional $F$. $\lambda[[F]]$ is assumed to be a functional itself, that is, $\lambda$ maps functionals to functionals  

Next, I introduce in addition to $P_2$ and $P_3$ two new projectors, $P_0$ and $P_1$:
\begin{eqnarray}
\label{P0P1}
P_0F[\phi]&\equiv& F[0],\\
P_1F[\phi]&\equiv& \big(\delta_\phi F[0]\big)\phi.
\end{eqnarray}
That is, $P_0$ separates the constant term in $F$, while $P_1$ separates the term linear in the field.

At last, using the above notations, I can write down conditions on the bare action of the Standard Model:
\begin{equation}
\label{condition}
(1-P_0-P_1-P_2-P_3)S_B=\lambda[[(P_1+P_2+P_3)S_B]].
\end{equation}
This equation states that the quartic self-couplings in the bare action are expressible it terms of the quadratic and cubic self-couplings. As we discused previously, the $\lambda$ in the right-hand-side can be expanded in powers of $P_3S_B$, and the expansions starts from the second power:
\begin{equation}
\label{expansion}
\lambda[[(P_2+P_3)S_B]]=\mathcal{O}[[(P_3S_B)^2]].
\end{equation}

In the next Section I will demonstrate that conditions (\ref{condition}) and (\ref{expansion}) suffice to express $A$ in terms of  $D_L$ and $T_L$.

\section{Iterative solution of the inaction equation}

Substitution of the representation (\ref{DE}) into condition (\ref{condition}) gives an equation for the generating functional $A[\phi]$. I call it the inaction equation because it does not involve the action of the theory. It may be advantageous because all the objects involved in this equation are UV finite. 

I discuss in this Section an iterative solution to the inaction equation. The first observation is that $A=0$ solves the equation. Indeed, $S_B=Q_L$ in this case (free theory) and the inaction equation is clearly satisfied.

Next, I try to solve it in the linear approximation assuming $A$ to be small and keeping only terms linear in $A$ in the equation:
\begin{equation}
\label{linear}
(1-P_0-P_1-P_2-P_3)TA[\phi]=0.
\end{equation}

The projector $1-P_0-P_1-P_2-P_3$ will enter prominently the following consideration, and I introduce a notation for it: $1-P_0-P_1-P_2-P_3\equiv P$. So, the projector $P$ projects away all the parts of a functional it acts upon if they are not more than cubic in the fields and are proportional to local operators present in the bare action. 

For solving (\ref{linear}), I notice that $T$-operation present in it does not mix the kernel and the image of $P$. Indeed, if $A$ is in the kernel of $P$, $TA=A+\text{term linear in}$ $\phi$. So, $TA$ is also in the kernel.

So, in the linear approximation, solution to the inaction equation are in the kernel of $P$. If I recall the definition of $A$ (\ref{amplitudes}), I see that the all the solutions are cubic in the fields, $A[\phi]=P_3A[\phi]$. 

To continue, it is convenient to use $T^{-1}$, the inversion of $T$:
\begin{equation}
\label{inverse}
T^{-1}=\exp\big(\frac{1}{2}D_L\delta_\phi\phi\big).
\end{equation}
This $T^{-1}$ also leaves the kernel of $P$ intact. Moreover,
\begin{equation}
\label{sandwich}
PT^{-1}(1-P)=0.
\end{equation}
This means that the image of $(1-P)$ (that is, the kernel of $P$) is mapped by $T^{-1}$ to itself. I will use this soon.

What I got by this moment is that, in the linear approximation in $A$, $A=T_L$ (see the definition (\ref{t-part})). Now, it is natural to decompose $A$ into the part in the image of $P_3$, and the part in its kernel: $A=T_L+V$, where $P_3T_L=T_L,$ and $P_3V=0$.

Notice that, in the linear approximation, $T_L$ is arbitrary vector in the finite-dimensional image of $P_3$, and $V$ vanishes. Substituting the above decomposition of $A$ into the inaction equation, I obtain an equation for $V$:
\begin{equation}
\label{decompositon}
P\log T\exp\big(-T_L-V\big)=-\bar{\lambda}[[T_L+V]].
\end{equation}
Here $\bar{\lambda}[[T_L+V]]=\lambda[[(P_1+P_2+P_3)S_B]]$, and one needs to substitute the right-hand-side of (\ref{DE}) in the argument of $\lambda$ instead of $S_B$.

Now I act on both sides of this equation with the operation $PT^{-1}$, recall the property (\ref{sandwich}), and obtain
\begin{equation}
\label{making-s}
PT^{-1}\log T\exp\big(-T_L-V\big)=-PT^{-1}\bar{\lambda}[[T_L+V]].
\end{equation}
The meaning of this transformation is that now the part of the left-hand-side linear in $T_L+V$ has become $P(-T_L-V)=-V$

Now I separate the part linear in $T_L+V$ in the left-hand-side of this equation and move the rest to the right-hand-side:
\begin{equation}
\label{almost}
V=-P\Big(T_L+V+T^{-1}\log T\exp\big(-T_L-V\big)\Big)+PT^{-1}\bar{\lambda}[[T_L+V]].
\end{equation}
Notice that the right-hand side of this equation as a series in $T_L+V$ starts from the second power because of the property (\ref{expansion}). Therefore, I can use this equation to obtain iteratively the expansion of $V$ in powers of $T_L$.

Indeed, let me assume that $V$ is a function of $T_L$. Expanding this function in a power series, I start with $[V]_{(0)}=0$, which corresponds to the above observation that $A=0$ is a free theory (here and below the number subscript in parenthesis means that the quantity bearing the subscript is taken with $T_L^n$ accuracy, that it the powers $T_L^{p>n}$ are dropped). Next, I have
\begin{equation}
\label{iterations}
[V]_{(n)}=\Big[-P\Big(T_L+[V]_{(n-1)}+T^{-1}\log T\exp\big(-T_L-[V]_{(n-1)}\big)\Big)+PT^{-1}\bar{\lambda}[[T_L+[V]_{(n-1)}]]\Big]_{(n)}.
\end{equation}
Notice that in the right-hand-side I have only $[V]_{(n-1)}$, which is known from the previous iteration. For example, $[V]_{(1)}=0$, and
\begin{equation}
\label{v-2}
[V]_{(2)}=\Big[-P\Big(T_L+T^{-1}\log T\exp\big(-T_L\big)\Big)+PT^{-1}\bar{\lambda}[[T_L]]\Big]_{(2)},
\end{equation}
and so on.

It's time to take a deep breath and observe what has been achieved.

\section{Conclusion}

In Section 2, input parameters of a theory have been extracted from its 
Green's functions. They are coordinates in a finite-dimensional image of the projector $P_2+P_3$, and the theory is parametrized with a vector of this space. The theory is defined by fixing this space, and by function $\lambda$ appearing in the right-hand-side of (\ref{condition}). 

In Section 4, the part of $A$ in the image of $P$ (denoted as $V$) was expressed as a series in $T_L$; the partial sums of this series are defined in terms of $Q_L$ and $T_L$. Because the Green's functions of the theory can be easily constructed by $V$, the generating functional $W$ of the Green's functions has been defined in this Section as a series in the input parameters in $T_L$;
the coefficients of this series are functions of the parameters in $Q_L$. A term of a fixed order in $T_L$ is a sum of Feynman-like diagrams with the same number of vertexes $T_L$. Individual diagrams may be UV divergent, but the sum should be finite for a renormalizable theory. That is, UV counterterms are generated in the course of the iterations described in this Section. The counterterms cubic in the fields are generated by applying the projector $P$, which subtracts local parts of functionals cubic in the fields. The counterterms quartic in the fields are generated by the function $\bar\lambda$ of (\ref{almost}).

Extraction of the input parameters is to an extent arbitrary. It is defined as soon as the projectors $P_2$ and $P_3$ are defined. But theory fixes only the images of these projectors. Obviously, there are infinitely many projectors sharing one and the same finite-dimensional image in an infinite-dimensional space. The arbitrariness of the extraction can be parametrized. A particular parametrization with a set of field configurations has been demonstrated in Section 2. The resulting Green's functions should be independent of the parameters of the extraction. This fixes the dependence of the input parameters on the extraction procedure. Consideration of this sort is a source of a new renormalization group equation generalizing the Gell-Mann--Low scheme. It has been tested  \cite{Pivovarov:2009wa} on the example of $\phi^4$. 

The immediate next problem is to consider the new renormalization group equation for the Standard Model. As discussed previously~\cite{Pivovarov:2015tya}, such a consideration can influence the analysis of the naturalness problem. 

Less immediate, but, probably, even more interesting is the problem of finding extensions of the Standard Model. Here I suggest  to look for an extension of the function $\lambda$ of (\ref{condition}) to the arguments from a space of larger dimension which would still yield a renormalizable theory. The ambitious hope is that new physical phenomena may be described by such an extension of the Standard Model. 

\section{Acknowledgements}

I thank Victor Kim and Sergei Trunov for collaboration. I also thank Nikolai Krasnikov and Dmitry Gorbunov for helpful discussions.

\end{document}